# Data assimilation of dead fuel moisture observations from remote automated weather stations

Martin Vejmelka[A,B,D], Adam K. Kochanski[C], and Jan Mandel[A,B]

[A]Department of Mathematical and Statistical Sciences, Campus Box 170, University of Colorado Denver, Denver, CO 80317-3364, USA

[B]Institute of Computer Science, Academy of Sciences of the Czech Republic, Prague, Czech Republic

[C]Department of Atmospheric Sciences, University of Utah, Salt Lake City, UT, USA

[D] Corresponding author. E-mail: vejmelkam@gmail.com

**Abstract:** Fuel moisture has a major influence on the behavior of wildland fires and is an important underlying factor in fire risk assessment. We propose a method to assimilate dead fuel moisture content observations from remote automated weather stations (RAWS) into a time-lag fuel moisture model. RAWS are spatially sparse and a mechanism is needed to estimate fuel moisture content at locations potentially distant from observational stations. This is arranged using a trend surface model (TSM), which allows us to account for the effects of topography and atmospheric state on the spatial variability of fuel moisture content. At each location of interest, the TSM provides a pseudo-observation, which is assimilated via Kalman filtering. The method is tested with the time-lag fuel moisture model in the coupled weather-fire code WRF-SFIRE on 10-hr fuel moisture content observations from Colorado RAWS in 2013. We show using leave-one-out testing that the TSM compares favorably with inverse squared distance interpolation as used in the Wildland Fire Assessment System. Finally, we demonstrate that the data assimilation method is able to improve fuel moisture content estimates in unobserved fuel classes.

**Introduction**

The behavior of fire is highly sensitive to fuel moisture content (FMC), which is defined as the mass of water per unit oven-dry mass of fuel. Fuel moisture affects the burning process in at least three ways (Nelson 2001): it increases ignition time, decreases fuel consumption and increases particle residence time. With increasing fuel moisture content, the spread rate decreases, and eventually, at the extinction moisture level, the fire does not propagate at all (Pyne *et al*. 1996). The moisture content depends on fuel properties and on atmospheric conditions. The fuel moisture content of live fuels exhibits predominantly a seasonal variation driven by physiological regulatory processes. In contrast, the fuel moisture content of dead fuels is influenced by a variety of weather phenomena such as precipitation, relative humidity, temperature, wind conditions, solar radiation and dew formation. For a recent review on modeling processes affecting fuel moisture in dead fuels, see Matthews (2014).

This paper reports on a method to assimilate dead fuel moisture observations supplied by sparsely situated remote automated weather stations (RAWS). An important feature of the method is the propagation of fuel moisture updates (derived based on the observed fuel moisture observations) to other unobserved fuel classes via an adjustment of common factors affecting fuel moisture content evolution in the model.

The method is presented in conjunction with the dead fuel moisture model implemented in WRF-SFIRE. The WRF-SFIRE model (Mandel *et al.* 2011) couples an established model of the atmosphere (Weather Research Forecasting model, WRF) (Skamarock *et al.* 2008), together with a model simulating fire behavior (spread fire model, SFIRE). The two components are connected via physical feedbacks — local wind speed drives the fire propagation, while the fire-emitted heat and vapor fluxes enter the weather model and perturb the state of the atmosphere in the vicinity of the fire. WRF-SFIRE has evolved from the Coupled Atmosphere - Wildland Fire - Environment model (CAWFE) (Clark *et al.* 2004). Similar models include MesoNH-ForeFire (Filippi *et al.* 2011). Recently, the WRF-SFIRE code has been extended by a fuel moisture model and coupled with the emissions model in WRF-Chem (Kochanski *et al.* 2012; Mandel *et al.* 2012). The current code and documentation are available online[1]. A version from 2010 is distributed with the WRF release as WRF-Fire (Coen *et al.* 2012; OpenWFM 2012).

**Methods**

*The dead moisture model in WRF-SFIRE*

The dead fuel moisture model in WRF-SFIRE (Kochanski *et al.* 2012; Mandel *et al.* 2012) simulates the moisture content in idealized, homogeneous fuel classes. They are commonly referred to by their drying/wetting time lag as 1-hour, 10-hour and 100-hour fuel (Pyne *et al.* 1996). The moisture content of each fuel class $k$ is simulated independently by a first-order differential equation with time lag $T_k$. The solution of the differential equation asymptotically approaches an equilibrium fuel moisture content, which depends on atmospheric conditions (temperature and relative humidity) and on whether the fuel undergoes drying (approaches the equilibrium from above) or wetting (approaches the equilibrium from below). If the fuel moisture lies between the drying and the wetting equilibria, it does not change. The effect of rain is modeled by the same type of time-lag equation, with the time lag value dependent on the rain intensity.

---
[1] http://openwfm.org

Denote the fuel moisture content of the $k$-th idealized fuel species with time lag $T_k$ by $m_k$, stored as a dimensionless proportion of mass of water per mass of wood. The fuel moisture model runs on a coarse mesh, with the nodes located at the middle of the faces of the atmosphere model on the ground. The moisture values of the idealized fuel classes are then integrated at every node of the finer fire model mesh. The relative contributions from the idealized fuel classes for each fuel type are derived from the 1h, 10h and 100h fuel loads presented by (Albini 1976, Table 7). The integration provides the fuel moisture estimates for an actual fuel type in each fire model cell, which is used then in the fire spread computations.

The fuel moisture model is described mathematically by the ordinary differential equation

$$\frac{d}{dt}m_k(t) = \begin{cases} \frac{S - m_k(t)}{T_r}\left(1 - \exp\left(\frac{r_0 - r(t)}{r_k}\right)\right), & \text{if } r(t) > r_0 \text{ (soaking in rain)} \\ \frac{E_d(t) - m_k(t)}{T_k}, & \text{if } r(t) \leq r_0 \text{ and } m_k(t) > E_d(t) \\ \frac{E_w(t) - m_k(t)}{T_k}, & \text{if } r(t) \leq r_0 \text{ and } m_k(t) < E_w(t) \\ 0, & \text{if } r(t) \leq r_0 \text{ and } E_w(t) \leq m_k(t) \leq E_d(t), \end{cases}$$

where $E_d(t)$ is the drying equilibrium, $E_w(t)$ is the wetting equilibrium, $S$ is the rain saturation level, $r_0$ is the threshold rain intensity, $r(t)$ is the current rain intensity, $r_k$ is the saturation rain intensity, $T_k$ is the drying/wetting time lag, and $T_r$ is the asymptotic soaking time lag in a very high-intensity rain. The coefficients $T_k$, $r_k$ and $r_0$ can be specified for each idealized fuel class by the user. The equilibria $E_d(t)$ and $E_w(t)$ are computed from the WRF-simulated rain intensity, as well as the air temperature and specific humidity at 2 m above the ground, similarly as in the fine fuel moisture component of the Canadian fire danger rating model (Van Wagner and Pickett 1985). In particular, the difference between the equilibria, $E_d(t) - E_w(t) > 0$, is constant. The parameters $S$, $T_r$, $r_0$ and $r_k$ were identified to match the behavior of the fuel soaking in rain in Van Wagner and Pickett (1985). The differential equation is solved by a numerical method exact for any length of the time step for coefficients constant in time. This is important because it allows for fuel moisture modeling on a much larger time scale (larger time steps) than fire behavior modeling. The above model is an empirical approach along the lines of Byram (1963), which takes into account vapor exchange and precipitation processes, but not other factors like solar radiation or soil moisture.

The present method assimilates observations into the fuel moisture content $m_k(t)$ and adjusts the equilibria $E_d(t)$, $E_w(t)$ and $S$. Since the equilibria are computed from external meteorological quantities, a standard solution is to extend the state of the model to also contain perturbations of the equilibria. By adding the perturbations to the model, we obtain an extended dynamical system for the variables $m_k$, $\Delta E$, and $\Delta S$,

$$\frac{d}{dt} m_k(t) = \begin{cases} \frac{S^A(t) - m_k(t)}{T_r}\left(1 - \exp\left(\frac{r_0 - r(t)}{r_k}\right)\right), & \text{if } r(t) > r_0 \\ \frac{E_d^A(t) - m_k(t)}{T_k}, & \text{if } r(t) \leq r_0 \text{ and } m_k(t) > E_d^D(t) \\ \frac{E_w^A(t) - m_k(t)}{T_k}, & \text{if } r(t) \leq r_0 \text{ and } m_k(t) < E_w^D(t) \\ 0, & \text{if } r(t) \leq r_0 \text{ and } E_w^D(t) \leq m_k(t) \leq E_d^D(t), \end{cases}$$

$$\frac{d}{dt}\Delta E(t) = 0,$$

$$\frac{d}{dt}\Delta S(t) = 0,$$

where we substitute assimilated environmental variables (annotated by the superscript A) for the original variables,

$$S^A(t) = \max(S + \Delta S(t), 0),$$

$$E_d^A(t) = \max(E_d(t) + \Delta E(t), 0),$$

$$E_w^A(t) = \max(E_w(t) + \Delta E(t), 0).$$

We write the discretization of the extended model as

$$\boldsymbol{m}(t) = f(\boldsymbol{m}(t-1), E_d(t, t-1), E_w(t, t-1), r(t, t-1)),$$

where the extended fuel moisture model state is

$$\boldsymbol{m}(t) = [m_1(t), m_2(t), \ldots, m_k(t), \Delta E(t), \Delta S(t)]^T$$

and $E_d(t, t-1), E_w(t, t-1)$ are the averages of the drying and the wetting moisture equilibria at time $t$ and $t-1$, $r(t, t-1)$ is the rain intensity in the same time interval.

The introduction of the assimilated parameters $\Delta E$ and $\Delta S$, which affect all fuel moisture classes, transforms the isolated equations for each fuel class into a coupled system. Such a coupling must be identified in any model, in which data assimilation is to indirectly affect the fuel moisture in unobserved fuel classes. Here, it is the equilibrium moisture content (modified via $\Delta E$ adjusted from the observed FMC), which affects the evolution of other fuel classes.

*Fitting model parameters to the domain of interest*

One set of the fuel moisture model parameters such as rain saturation level ($S$), the drying/wetting time lag ($T_k$), and the asymptotic soaking time lag ($T_r$), may be not optimal for all environments. Therefore in this work, we first searched for an optimal set of these parameters appropriate for the State of Colorado, by fitting the fuel moisture model to the past data using a grid search optimization procedure. We have retrieved 10-hr fuel moisture, air temperature, relative humidity and accumulated precipitation data from all 45 stations that provided 10-hr fuel moisture observations in 2012 and from all 30 stations that provided observations in 2013. We then discretized the parameter space for each of the parameters $S$ (in steps of 0.2), $T_r$ (in steps of 1 hr), $r_0$ (in steps of 0.01 mm/h), and $r_k$ (in steps of 1 mm/h) and $\Delta E$ (in steps of 0.01) and simulated the dead fuel moisture for all stations for the entire year, using all possible parameter combinations. Then the model results were compared with observations in order to select the set of parameters minimizing the mean squared error in the 10-hr fuel moisture estimate. Each fuel moisture model run was initialized at its first data point of the year using the average of the drying and wetting equilibria. While this procedure is computationally intensive, it does not need to be done often and guarantees a global optimum at the given resolution. Table 1 summarizes the results of this fitting:

|  | $S$ [−] | $T_r$ [hr] | $r_0$ [mm/h] | $r_k$ [mm/h] | $\Delta E$ [-] |
|---|---|---|---|---|---|
| Kochanski *et al.*, 2012 | 2.5 | 14 | 0.05 | 8 | 0 |
| Search space | 0.2 – 2.4 | 4 – 16 | 0.01 – 0.12 | 1 – 12 | -0.1 – 0.1 |
| Colorado (2012) | 0.6 | 7 | 0.08 | 2 | -0.04 |
| Colorado (2013) | 0.4 | 6 | 0.1 | 1 | -0.04 |

**Table 1 Original and optimized parameters of the fuel moisture model to Colorado remote automated weather station observations. The parameters have been fitted to observations in the year 2012 and in 2013 separately.**

As shown in Table 1, the optimized fuel model parameters derived from 2012 and 2013 are quite similar, especially compared to the range of possible values and the discretization of the parameter space. The negative values of $\Delta E$ suggest that the fuel moisture equilibria were generally overestimated when the original set of fuel parameters was used for the State of Colorado.

Table 2 shows the error statistics for the original fuel moisture runs with default parameters, and the new runs with the fuel moisture parameters optimized based on the 2012 observational data. The parameter optimization significantly reduced the mean errors (bias) in the fuel moisture estimates for both analyzed years as well as the mean absolute errors.

| Parameters/year | Mean error (bias) [-] | Mean abs. error [-] | Corr. coeff. [-] |
| --- | --- | --- | --- |
| Original/2012 | 0.047 | 0.053 | 0.75 |
| Original/2013 | 0.058 | 0.063 | 0.70 |
| Optimized/2012 | 0.003 | 0.030 | 0.75 |
| Optimized/2013 | 0.016 | 0.034 | 0.74 |

**Table 2 Errors of fuel moisture model for original parameters fitted to data of Van Wagner and Pickett (1985) and with parameters optimized for RAWS observations in 2012. Note that the row 'Optimized/2012' contains statistics on the fitted data (in-sample error), while 'Optimized/2013' is an out-of-sample error.**

An example of the effect of the parameter optimization on the simulated 10-hr fuel moisture content is presented in Figure 1. The parameter adjustment assures a better general agreement between the simulations and observations.

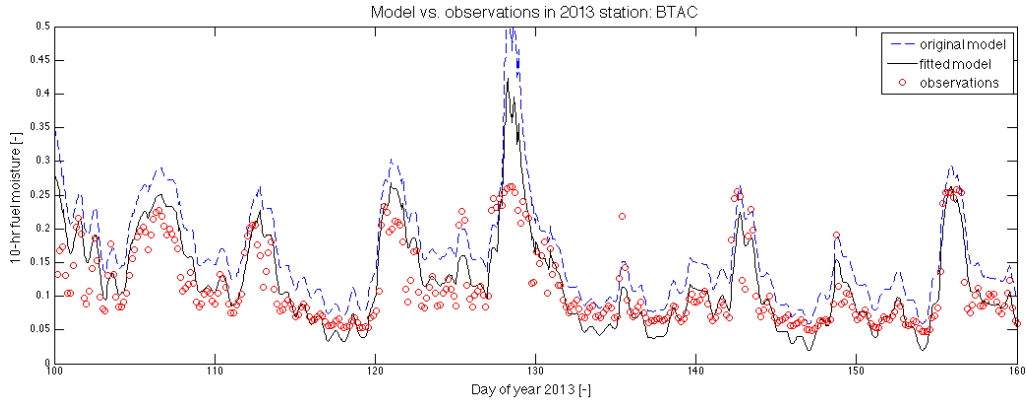

**Figure 1** Model trace for station BTAC2 (Sugarloaf: 40.018N, 105.361W) with original parameters, parameters fitted to Colorado station data 2012 compared here to station observations for days 100 to 160 of year 2013.

*Kalman filtering*

Data assimilation is a mechanism for combining observations with model forecasts to produce more accurate estimates of state or model parameters. Kalman filtering is a frequently used method of data assimilation in many areas of engineering (Simon 2010).

In the previous section, we have optimized the one set of fuel moisture parameters for the entire State of Colorado in order to obtain a good starting estimate of the behavior of 10-hr dead fuel moisture. However, Kalman filtering can adjust the fuel moisture model parameters on a location-specific basis reflecting local environmental conditions.

For the purpose of Kalman filtering, we restate the discrete version of the model in a stochastic setting as

$$\boldsymbol{m}(t) = f\big(\boldsymbol{m}(t-1), E_d(t,t-1), E_w(t,t-1), r(t,t-1)\big) + \boldsymbol{w}(t-1),$$

$$y(t) = h(\boldsymbol{m}(t)) + \boldsymbol{v}(t),$$

where $\boldsymbol{w}(t)$ is the process noise, which represents the growth in uncertainty of the state due to imperfections in the model, $\boldsymbol{v}(t)$ is the observation noise, and $y(t)$ represents the observations predicted by the model. Both noise terms are assumed to be zero-mean, uncorrelated and white. The covariance of the process noise is a parameter of the Kalman filter and it will be denoted by $Q$, while the observation covariance $R$ is provided together with each observation.

We considered and compared two variants of Kalman Filtering (KF), the Extended Kalman Filter (EKF) (Simon 2010, §13.2.3) and the Unscented Kalman Filter (UKF) (Julier and Uhlmann 1997, 2004; Julier *et al*. 2000) to assimilate the pseudo-observations at each grid point as supplied by the trend surface model presented in the next section.

At each time step, a Kalman filter executes in two phases. First, the last moisture state $\boldsymbol{m}(t-1)$ is evolved into the forecast $\widehat{\boldsymbol{m}}(t)$ and the covariance $P(t-1)$ is propagated to the forecast covariance $\widehat{P}(t)$. In the second phase, if an observation is available, the forecasts are updated to the analysis $\boldsymbol{m}(t)$ and $P(t)$. If an observation is not available, then the forecast values become the new analysis values.

Kalman filters must be initialized with a mean $\boldsymbol{m}(0) = \boldsymbol{m_0}$ and background covariance $P(0) = P_0$. In all our experiments, the initial state mean is set to the average of the drying and wetting equilibrium at $t_0$. The background covariance $P_0$ is diagonal and is set to 0.01 for each fuel class FMC and to 0.001 for each fuel moisture parameter $(\Delta E, \Delta S)$, since these are fitted to larger sets of data. We now discuss the two types of Kalman filters tested.

The Extended Kalman filter models the evolution of the fuel moisture by passing the current estimate of the state through the discretized model function

$$\widehat{\boldsymbol{m}}(t) = f\big(\boldsymbol{m}(t-1), E_d(t, t-1), E_w(t, t-1), r(t, t-1)\big).$$

The forecast covariance is computed using the Jacobian $J_f$ of the model as

$$\hat{P}(t) = J_f(t-1)P(t-1)J_f(t-1)^T + Q.$$

The term $J_f P_{t-1} J_f^T$ is equal to the first term in the Taylor expansion of the exact covariance propagation through the nonlinear function $f$. The EKF thus has first order accuracy in covariance propagation, as higher order terms in the Taylor expansion are missing.

The update phase can be summarized as

$$K(t) = \hat{P}(t)H\left(H\hat{P}(t)H^T + R(t)\right)^{-1},$$

$$\boldsymbol{m}(t) = \hat{\boldsymbol{m}}(t) + K(d(t) - H\hat{\boldsymbol{m}}(t)),$$

$$P(t) = (I - K(t)H)\hat{P}(t),$$

where $K(t)$ is the Kalman gain at time $t$, $d(t)$ is the observation and $R(t)$ is the covariance of the observation. $H$ is the observation operator, which has a particularly simple form for our problem as the 10-hr fuel moisture is observed directly.

The Unscented Kalman filter takes a different approach. It is based on the unscented transformation, which is a deterministic sampling technique for propagating the statistics of a random variable through a nonlinear transformation (Julier and Uhlmann 1997). For an $n$-dimensional random variable $x$, $2n + 1$ *sigma points* are selected in the state space so that the mean and covariance of the state are equal to the sample covariance and the sample mean of the sigma points. Each of these points is passed through the nonlinear model function and the new mean and covariance are set to the sample mean and sample covariance of the propagated sigma points. The sigma points are chosen so that the forecast covariance matches the covariance of the propagated covariance at least to the second term using the Taylor expansion of the model function, and thus it has second-order accuracy in the small variance asymptotics (Julier and Uhlmann, 2004), whereas the EKF, as a linearization, only has first-order accuracy. For details on the UKF procedure, see Julier and Uhlmann (2004).

Since it requires multiple evaluations of the model, the Unscented Kalman filter is typically more computationally intensive than the Extended Kalman filter. On the other hand, it does not require one to compute the Jacobian, making it easier to use existing fuel moisture model codes.

*Estimating the fuel moisture field from sparse surface observations*

Surface observations are generally sparse and provide the dead fuel moisture only at the locations of the measurement stations. Without additional processing this observational dataset does not provide information on the fuel moisture for other locations (between the observational stations). In particular the discrete observations are not suitable for spatial initialization of the fuel moisture in the fire spread models, requiring a gridded data set providing the fuel moisture estimate at each model grid point. In order to remove this limitation, we develop a mechanism to estimate the fuel moisture at arbitrary points based on the available observations from other remote automated weather stations (RAWS). This approach has been applied in order to estimate the fuel moisture at each grid point in our test domain. We assume that the evolution of the dead fuel moisture is affected by the local topography and the atmospheric state. It is therefore natural to use local such variables to model the spatial variability of fuel moisture. We fit a linear regression model using such predictors to all observations valid at a given time and the estimated coefficients are then used to supply pseudo-observations and their estimated variances at each location.

We use a variant of the trend surface modeling approach proposed by (Schabenberger and Gotway 2005, §5.3.1), which is mathematically equivalent to a model introduced by Fay and Herriot (1979) to compute estimates of income for small areas based on census data. On the regional scale, we prefer this method to a full universal kriging approach and argue our viewpoint in the discussion section. The assumed form of fuel moisture observation $Z(s)$ at location $s$ is

$$Z(s) = \beta_1 X_1(s) + \cdots + \beta_k X_k(s) + e(s) = x(s)\beta + e(s),$$

where the predictor fields $X_i$, also called covariates, are known at every location $s$, $\beta_i$ are unknown regression coefficients, the error $e(s)$ is independent at each grid point and $x(s) = [X_1(s), X_2(s), \ldots, X_k(s)]$ is the row vector of covariates at an arbitrary location $s$. The error $e(s)$ is assumed to have zero mean and consist of an independent observation error with variance $\gamma^2(s)$ assumed known, and a microscale variability with variance $\sigma^2$, which is unknown but constant in the domain (Cressie 1993). In our model, the microscale variability additionally captures the errors incurred by the linear regression model itself due to having fewer covariates than observations, which is the standard situation. Microscale variability reflects subgrid-scale effects that cannot be adequately captured at the spatial resolution of the model. We write this observation model in a compact matrix form for all locations of interest simultaneously,

$$Z = X\beta + e, \quad e \sim \mathcal{N}(0, \Sigma), \quad \Sigma = \Gamma + \sigma^2 I,$$

where $\Gamma = \text{diag}(\gamma^2(s))$ and $X = [X_1, X_2, \ldots, X_k]$ is the matrix of regressors.

The coefficients $\beta$ and the microscale variability variance $\sigma^2$ are estimated from the data at every time step. Given the microscale variability variance $\sigma^2$, observations $Z$ and covariates $X$ at the same locations $s = [s_1, s_2, \ldots, s_n]$, the standard least-squares estimate $\hat{\beta}$ of the regression coefficients is

$$\hat{\beta} = (X^T \Sigma^{-1} X)^{-1} X^T \Sigma^{-1} Z,$$

where $\Sigma$ is the covariance matrix corresponding to the locations of the observations. To estimate the microscale variability variance $\sigma^2$, we numerically solve the equation

$$\sum_{i=1}^{n} \frac{\hat{e}(s_i)^2}{\gamma^2(s_i) + \hat{\sigma}^2} = n - k,$$

for $\hat{\sigma}^2$ where $\hat{e}(s_i) = Z(s_i) - x(s_i)\widehat{\beta}$ are the residuals at location $s_i$, $n$ is the number of locations observed and $k$ is the number of regressors (Fay and Herriot 1979). Both estimates $\widehat{\beta}$ and $\hat{\sigma}^2$ are found by an iterative method starting from $\hat{\sigma}^2 = 0$. In each iteration, the method first estimates $\widehat{\beta}$ and then $\hat{\sigma}^2$ until convergence.

The Kalman filter at location $s$ then receives a pseudo-observation $d(s) = x(s)\widehat{\beta}$, which is assigned the variance

$$R(s) = \hat{\sigma}^2 + x(s)(X^T \Sigma^{-1} X)^{-1} x^T(s).$$

The derivation of the pseudo-observation variance $R(s)$ can be found in the Appendix.

Finally note that due to the nature of the trend surface model, it is possible that for some locations, negative values of fuel moisture content are predicted. These must be trimmed to 0 to prevent the appearance of negative fuel moisture content in the fuel models.

*RAWS 10-hr fuel moisture observations*

Some Remote Automated Weather Stations (RAWS) have 10-hr fuel stick sensors and provide hourly measurements of fuel moisture content (FMC). We obtain these observations and metadata from the MesoWest[2] website. The FMC observations are provided as the number of grams of water in 100g of wood. Before assimilation, these are rescaled to a dimensionless value in the range 0 to 1, in order to match the representation of the FMC in the model.

---
[2] http://mesowest.utah.edu/

Unfortunately, information on the type of fuel moisture sensors fitted to each station is unavailable in the MesoWest network. As a rough guideline, we have used the manual (Campbell Scientific, 2012) for the fuel stick sensor CS-506 from Campbell Scientific to assign a variance to all RAWS observations. We note that the variance of the observation depends on the fuel moisture content, thus necessitating the trend surface model with unequal variances of observations at different locations.

**Results**

*Leave-one-out testing in Colorado with 2013 observations*

We first perform detailed tests of the trend surface model approach using 2013 station data in Colorado. Use of station data allows us to avoid the impacts of the weather forecast accuracy and the representation errors associated with the model spatial grid not collocated with the locations of the observational stations. We perform all tests using parameters of the moisture model optimized for 2012 Colorado RAWS observations and run all tests using observations from Colorado stations collected in year 2013.

We run three variants of the trend surface model. In the first variant, the trend surface model is used with four covariates: station elevation, a constant term, rain intensity and the atmospheric moisture equilibrium computed from station relative humidity and air temperature. This variant requires no fuel moisture model and is suitable simply for spatial extrapolation of fuel moisture observations. The rain intensity covariate is removed if there was no rain over the domain to prevent the appearance of singular matrices. This variant will be denoted 'TSM'.

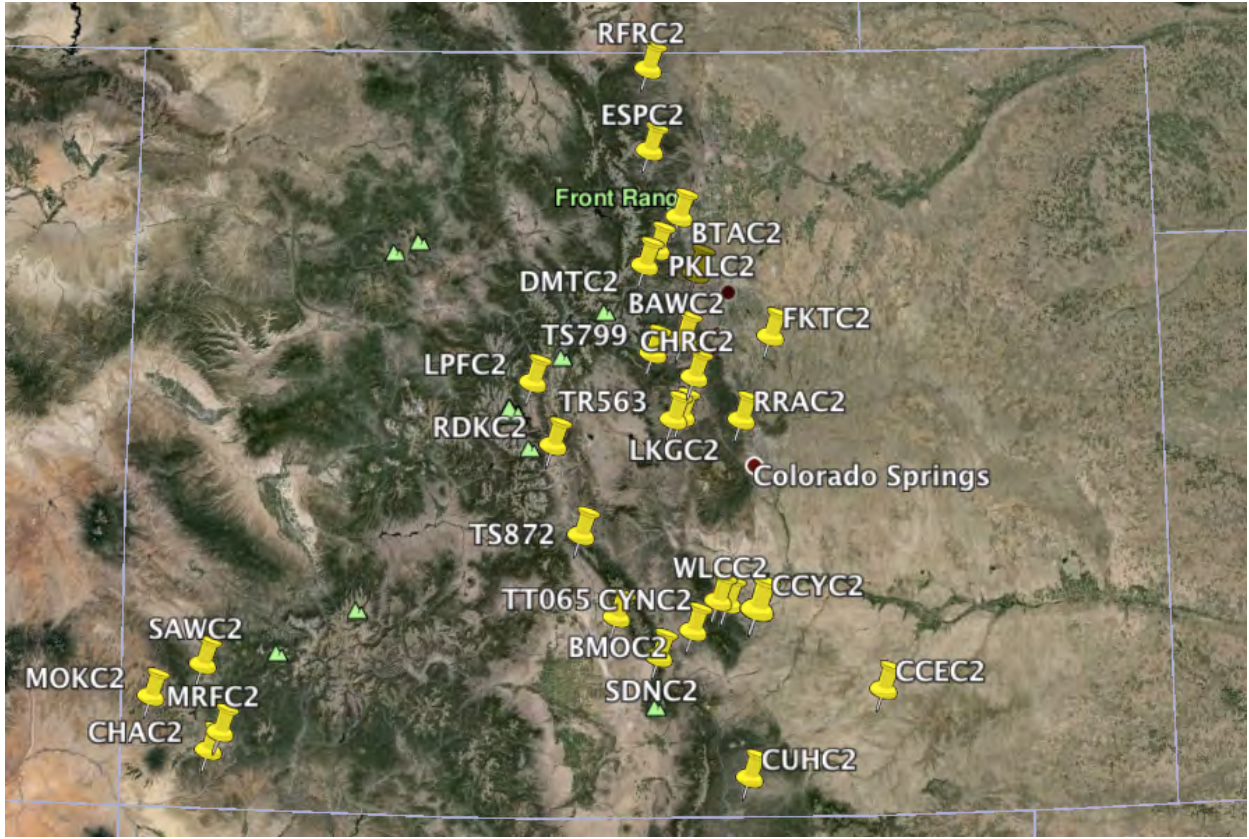

**Figure 2. The 28 Colorado remote automated weather stations that have supplied 10-hr fuel moisture observations in 2013 (image from Google Earth, station locations from University of Utah MesoWest network).**

In the second variant, the atmospheric moisture equilibrium is replaced by the forecast of a fuel moisture model running at the location of each transmitting station. The Extended Kalman Filter is used to assimilate the pseudo-observations provided by the trend surface model into the fuel moisture model thus constructing a coupled system where the fuel moisture models provide spatial structure to the trend surface model, which in turn provides the next pseudo-observations. This variant will be denoted 'TSM+EKF'.

The third variant is similar to the second variant, where we replace the Extended Kalman Filter by the Unscented Kalman Filter. This variant will be denoted 'TSM+UKF'.

We shall compare these variants to the inverse square distance interpolation method currently used in the Wildland Fire Assessment System (Burgan *et al.* 1998). This method is denoted 'INTERP2'.

To estimate the error incurred by each of the above methods, we turn to leave-one-out testing. For each of the 28 stations (see Figure 2 for locations), we leave all of its 10-hr fuel moisture observations out and attempt to predict them using the remaining data (i.e. including weather conditions at the left-out station) at each time point. We note that leave-one-out testing provides an unbiased estimate of the prediction error.

The results of this test are summarized in Figure 3 as mean absolute prediction errors (MAPE) for each left-out station. All methods are also compared to running the WRF-SFIRE fuel moisture model, denoted by 'MODEL', based on station temperature and relative humidity observations only. This serves as a benchmark for the remaining methods. The total number of 10-hr fuel moisture observations involved in this test is 210503.

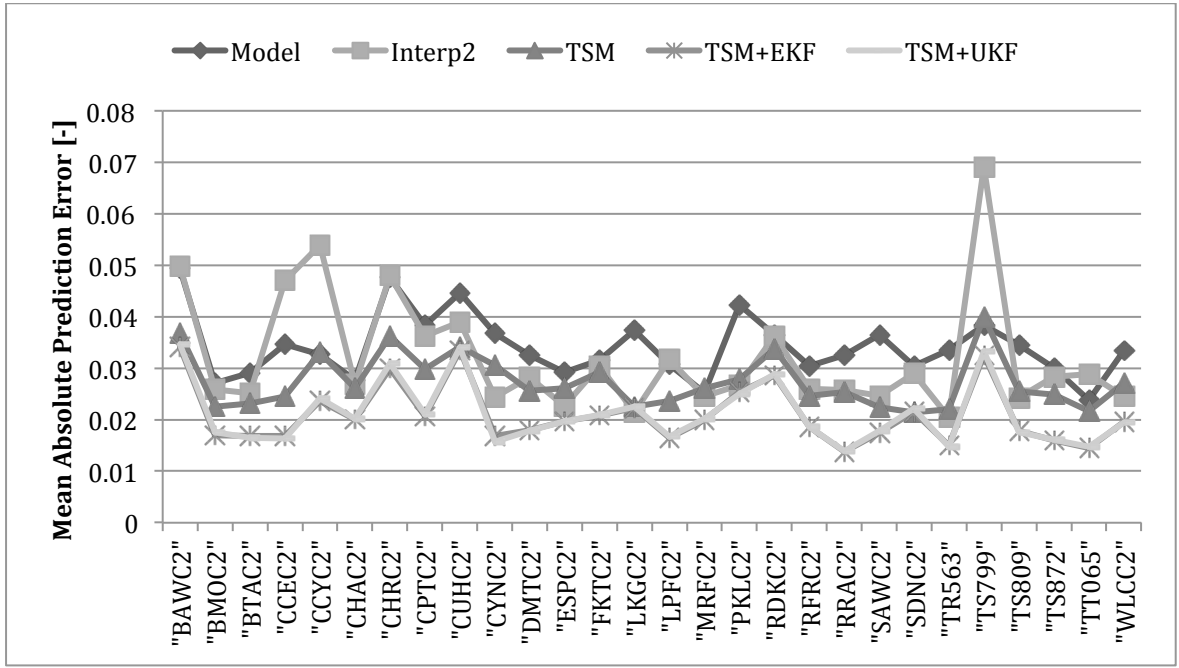

**Figure 3 Mean absolute prediction errors for each station from the leave-one-out test using RAWS 10-hr fuel moisture observations from 2013 in Colorado.**

A few features are apparent in the above plot, for example the UKF and EKF have performed very similarly. This was expected to some extent and has led us to choose the UKF filter over the EKF filter for reasons relegated to the discussion section. Clearly, the use of Kalman filtering coupled with the trend surface model worked best for almost all of the stations. We attribute the improvement over using the fuel moisture equilibrium as a covariate to the fact that 10-hr fuel moisture is often quite far from the moisture equilibrium, so the current fuel moisture forecast captures the structure of the fuel moisture field better than the atmospheric equilibrium.

Table 3 gives the mean average prediction errors over all stations and all 10-hr fuel moisture observations in 2013 in Colorado for each method and the relative improvement over the fuel moisture model without data assimilation and over the inverse square interpolation method:

| Method | Model | Interp2 | TSM | TSM+EKF | TSM+UKF |
|---|---|---|---|---|---|
| **MAPE [-]** | 0.0342 | 0.0322 | 0.0274 | 0.0207 | 0.0209 |
| **vs. Model** | 0% | 5.85% | 19.88% | 39.47% | 38.89% |
| **vs. Interp2** | -6.21% | 0% | 14.91% | 35.71% | 35.09% |

Table 3. Leave-one-out error for different methods of estimating the 10-hr fuel moisture content field. The second and third rows give relative improvement with respect to using model with out data assimilation and with respect to using only the inverse squared distance interpolation.

The results of the experiment support the claim that the trend surface model is an improvement over the inverse square distance interpolation method when using only the station observations. However, coupling the trend surface model with the fuel moisture model and Kalman filtering brings yet more improvement toward a MAPE close to 0.02.

*Effect of data assimilation on unobserved fuels*

We investigate quantitatively the effect of data assimilation of 10-hr fuel moisture content observations on 1-hr and 100-hr fuel moisture. Unfortunately, 1-hr and 100-hr FMC is sampled sporadically, about once or twice per month at very few locations. We have obtained observations (collected at 2pm MDT) of 1-hr FMC from 8 stations (38 observations total, May – August 2013) and 100-hr FMC from 9 stations (51 observations total, May – August 2013) from the Wildland Fire Assessment System database[3]. These observations are not co-located with the RAWS locations and we must obtain the atmospheric state and precipitation fields from another source. We have opted to use the Real-Time Mesoscale Analysis (RTMA) provided by NOAA/NCEP, as it is available hourly at a 2.5km resolution for the entire contiguous United

---

[3] www.wfas.net

States (CONUS). We have additionally varied the data assimilation period (simulation run-time) to observe the effect of longer data assimilation and of the time of day when the fuel model is initialized.

Table 4 summarizes the impact of assimilating 10-hr FMC observations on the 1-hr and 100-hr fuels. Since there are few observations, we also supply standard errors of mean.

|  | Raw Mean Error [-] | DA Mean Error [-] | Raw MAPE [-] | DA MAPE [-] |
|---|---|---|---|---|
| 1-hr fuel/6 hours | -0.051 ± 0.009 | -0.021 ± 0.009 | 0.058 ± 0.007 | 0.039 ± 0.007 |
| 1-hr fuel/12 hours | -0.051 ± 0.009 | -0.024 ± 0.009 | 0.058 ± 0.007 | 0.042 ± 0.007 |
| 1-hr fuel/24 hours | -0.051 ± 0.009 | -0.020 ± 0.009 | 0.058 ± 0.007 | 0.040 ± 0.007 |
| 1-hr fuel/48 hours | -0.051 ± 0.009 | -0.017 ± 0.009 | 0.058 ± 0.007 | 0.040 ± 0.007 |
| 100-hr/6 hours | 0.031 ± 0.008 | 0.032 ± 0.008 | 0.050 ± 0.006 | 0.051 ± 0.006 |
| 100-hr/12 hours | 0.059 ± 0.010 | 0.060 ± 0.010 | 0.070 ± 0.008 | 0.070 ± 0.008 |
| 100-hr/24 hours | -0.009 ± 0.010 | -0.006 ± 0.010 | 0.048 ± 0.006 | 0.047 ± 0.007 |
| 100-hr/48 hours | 0.012 ± 0.011 | 0.015 ± 0.011 | 0.055 ± 0.008 | 0.054 ± 0.008 |

**Table 4 Fuel moisture content prediction errors (with standard errors of mean) for model without data assimilation and model with data assimilation for different unobserved fuel types and different simulation times.**

While the small number of observations is not conducive to rigorous statistical testing, some differences appear large enough for interpretation in this exploratory analysis. We summarize the findings in that data assimilation on the tested time scales improves the estimates of the 1-hr FMC by about 30% in terms of the MAPE and by about 60% in terms of the mean error. However, no discernible improvement is visible in the 100-hr FMC apart from the effect of the longer simulation run (not attributable to the data assimilation). Since 100-hr fuel has a time constant of approximately 4 days, longer data assimilation runs may be needed to observe an improvement. We also note that initializing the fuel model at an inopportune time causes the model to incur higher errors in the slow 100-hr fuel. In our results, this is visible for the 12-hour runs, where the fuel moisture is initialized at 4am in the local time zone close to the peak equilibrium fuel moisture content, while the 6-hour runs (initialized from fuel moisture equilibrium at 10am) show smaller errors even though the run is shorter. An example of the 1h fuel moisture estimated for Colorado for 6/11/2013 at 2pm MDT using the TSM+UKF method is presented in Figure 4.

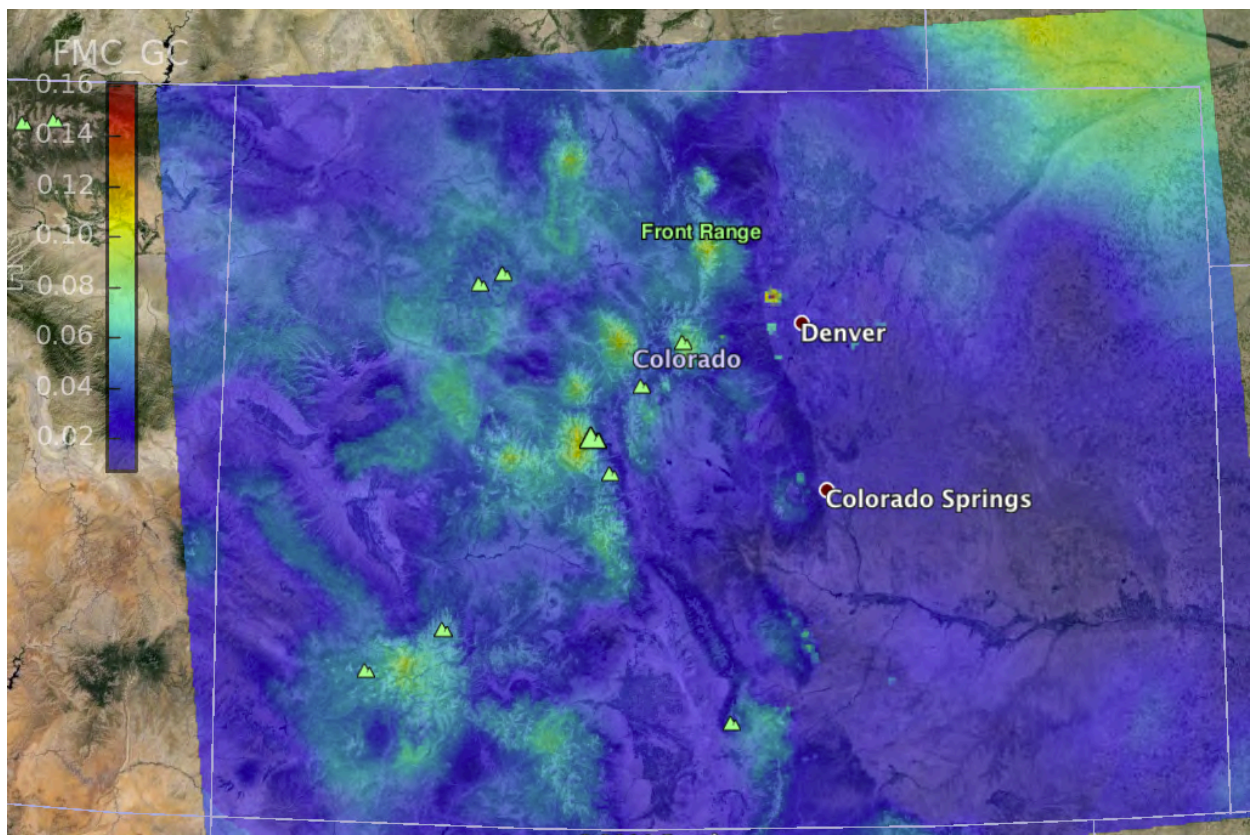

**Figure 4 Fuel moisture field (1-hr fuel, 6/11/2013 2pm MDT) generated using the RTMA to supply atmospheric state.**

**Discussion**

*Unscented vs. Extended Kalman filter*

The Extended Kalman filter has been the mainstay of data assimilation involving nonlinear models (Simon 2010; Julier and Uhlmann 2004). In our experiments, we did not find a substantial difference in performance between that the Extended Kalman filter and the Unscented Kalman Filter. However, in our final design, we have opted for the UKF for several reasons. Its performance is likely to be good even in situations with longer intervals between fuel moisture updates, as its propagation of forecast covariance is accurate to the second order, whereas the EKF uses a first-order approximation valid at the start of the integration interval. The UKF also has the very important advantage that it does not require a Jacobian for covariance propagation, making it easier to reuse the data assimilation mechanism with new fuel moisture models. In our experiments, the UKF performed as well as the more established EKF and offers significant implementation advantages, thus motivating our decision to use it.

*Trend surface model compared to other strategies*

In Burgan (1998), the authors noted that their inverse squared distance interpolation strategy does not account for the effects of topography or atmospheric state variability. The trend surface model is able to explicitly use topography and atmospheric state as auxiliary information and moreover provides a measure of uncertainty of the pseudo-observations computed at each time point and location.

The objective of the proposed method is its integration in a routinely used fuel moisture assimilation mechanism. Strong emphasis on the stability of the numerical algorithms is thus important in addition to minimal user intervention requirements. An alternative to the trend surface model is universal kriging, which attempts to leverage spatial correlations in model errors by specifying a covariance model. In complex terrain, a complicated model of covariance would be necessary to exploit any residual spatial relationships in the model errors. An examination of variograms of fuel moisture observations in the Front Range region of Colorado has not uncovered a convincing distance-related structure. We also note that universal kriging is typically used in much smaller or much larger domains, at scales where assumptions on smoothness of the topography and atmospheric forcing facilitate the construction of distance-based models of covariance, while at the mesoscale level, non-stationarity induced by weather phenomena and terrain properties makes use of universal kriging methods challenging in the least.

We also note that the trend surface model approach is highly extensible. If a new source of spatial data relevant to fuel moisture content becomes available (e.g. a high resolution soil moisture product), it can be objectively tested for its predictive power using the leave-one-out strategy we have already used in our work. If the new field reduces the leave-one-out error then it can be incorporated in the algorithm as another predictor.

A more detailed approach would take into account the uncertainty in model-generated covariates, which are themselves loaded with errors. We have considered using the total least squares framework but this formulation has a condition that is always worse than that of the standard least squares problem (Golub and Van Loan, 1980). The question whether this would improve the performance of the data assimilation system is still open.

*Future developments*

In future research, we will concentrate on use of remote sensing products to provide additional predictor variables for the trend surface model. The TSM could also be improved by using a constraint optimization algorithm that would prevent the appearance of negative values that now must be culled to zero. We would also like to integrate this framework with a weather model to perform forecasting of dead fuel moisture.

**Conclusion**

The objective of the reported work was to provide improved fuel moisture content estimates for fire behavior modeling in on-demand fire modeling scenarios and to systems for operational fire risk estimation. With this in mind, we have proposed a computationally efficient and extensible method for the assimilation of point dead fuel moisture observations into fuel moisture models. The method has been tested in conjunction with the fuel moisture model used in WRF-SFIRE and 10-hr fuel moisture observations from Remote automated weather stations.. We have demonstrated using leave-one-out testing that the proposed method is able to capture the spatial variability of the fuel moisture field and reduce the absolute error of the estimates of the observed fuel by about 40% compared to running an optimized fuel moisture model and by about 35% compared to inverse squared distance interpolation. Further numerical experiments have shown that data assimilation also improves estimates of the unobserved 1-hr fuel moisture content, while longer data assimilation runs may be needed to improve estimates of 100-hr fuel moisture content.

**Acknowledgments**

This research was partially supported by the National Science Foundation (NSF) grants AGS-0835579 and DMS-1216481, National Aeronautics and Space Administration (NASA) grants NNX12AQ85G and NNX13AH9G, and the Grant Agency of the Czech Republic grant 13-34856S. The authors would like to thank the Center for Computational Mathematics University of Colorado Denver for the use of the Colibri cluster, which was supported by NSF award CNS-0958354. This work partially utilized the Janus supercomputer, supported by the NSF grant CNS-0821794, the University of Colorado Boulder, University of Colorado Denver, and National Center for Atmospheric Research.

*Appendix*

To derive the equation for the variance of the pseudo-observation, we split the equation

$$Z = X\boldsymbol{\beta} + \boldsymbol{e}, \boldsymbol{e} \sim \mathcal{N}(0, \Sigma), \Sigma = \Gamma + \sigma^2 I,$$

into two parts and write it for a single location. The first part models the underlying true fuel moisture field:

$$S = x\boldsymbol{\beta} + \eta, \eta \sim \mathcal{N}(0, \sigma^2),$$

where $\eta$ is the microscale variability and $\gamma^2$ is its variance. The second is the observation model:

$$Z = S + \epsilon, \epsilon \sim \mathcal{N}(0, \gamma^2),$$

where $\epsilon$ is the measurement error and $\sigma^2$ is its variance. Our problem is to estimate the variance of the true underlying fuel moisture field $S$ given the least square estimate

$$\widehat{\boldsymbol{\beta}} = (X^T \Sigma^{-1} X)^{-1} X^T \Sigma^{-1} \mathbf{Z}.$$

We have that

$$\text{var}(S) = \text{var}(x\widehat{\boldsymbol{\beta}} + \eta) = \text{var}(x\widehat{\boldsymbol{\beta}}) + \sigma^2,$$

and that

$$\text{var}(x\widehat{\boldsymbol{\beta}}) = x(X^T \Sigma^{-1} X)^{-1} x^T,$$

so that finally

$$\text{var}(S) = \sigma^2 + x(X^T \Sigma^{-1} X)^{-1} x^T.$$

The last step is to substitute our estimates to obtain the estimate of the variance for the pseudo-observation at location $s$

$$R(s) = \hat{\sigma}^2 + x(s)\left(X^T \widehat{\Sigma}^{-1} X\right)^{-1} x(s)^T.$$

Where we note that $\widehat{\Sigma} = \Gamma + \hat{\sigma}^2 I$, so the estimated microscale variability variance appears also in the second term of the expression.